# Síntesis de películas delgadas de ZnS por CBD para uso como capa *buffer* en celdas solares basadas en $Cu_2SnZnS_4$

Saul Daniel Cruz Lemus[1*], John Michael Correa Hoyos[1**], William Chamorro Coral[1***], Raúl Andrés Becerra Arciniegas[1], Hugo Alexander Suarez[1], Oscar Rodríguez Bejarano[1], Gerardo Gordillo Guzmán[2]

[1] Departamento de Química. Universidad Nacional de Colombia. Bogotá - Colombia
[2] Departamento de Física. Universidad Nacional de Colombia. Bogotá - Colombia

**Resumen** Películas delgadas de ZnS fueron sintetizadas por el método CBD *(Chemical bath deposition)* usando una solución constituida por tiourea y acetato de zinc como fuentes de S y Zn y citrato de amonio como agente acomplejante. Estas fueron crecidas sobre sustratos de vidrio soda lime recubiertos con películas delgadas de $Cu_2SnZnS_4$ (CZTS). A través de múltiples ensayos realizados variando los principales parámetros de síntesis se encontraron condiciones para crecer películas delgadas de ZnS con propiedades adecuadas para ser usadas como capa *buffer* de celdas solares basadas en $Cu_2SnZnS_4$.

Las películas de ZnS fueron caracterizadas usando técnicas tales como espectrofotometría, Difracción de rayos-x (DRX) y microscopía de fuerza atómica (AFM). Medidas de DRX revelaron que las películas de ZnS con espesores menores que 200 nm, crecidas sobre vidrio y sobre $Cu_2SnZnS_4$ presentan una estructura amorfa; sin embargo, muestras con espesores mayores de 300 nm crecen con estructura cristalina correspondiente a la fase $ZnSO_8H_8$ (sulfohidróxido de zinc), indicando que la película formada es realmente una mezcla de ZnS e hidróxidos de Zn, conocidos genéricamente en la literatura como Zn(S,OH).

**Abstract** ZnS thin films were deposited by CBD (chemical bath deposition) method using a solution containing thiourea and zinc acetate as sources of S and Zn and ammonium citrate as complexing agent. The samples were grown on soda lime glass substrates covered with a $Cu_2SnZnS_4$ (CZTS) thin film. Through an extensive parameter study carried out varying the main deposition parameter in a wide range, conditions were found to grow ZnS thin films with suitable properties to be used as buffer layer in $Cu_2SnZnS_4$ based solar cells.

---

[*] sdcruzl@unal.edu.co
[**] jmcorreah@unal.edu.co
[***] williamchamorrocoral@gmail.com

The ZnS films were characterized using techniques such as spectrofotometry, X-ray diffraction (XRD) and atomic force microscopy (AFM). XRD measurements revealed that ZnS films with thicknesses lower than 200 nm grow with amorphous structure; however, samples with thicknesses higher than3 nm present crystalline structure corresponding to the $ZnSO_8H_8$ (zinc sulphohydroxide) phase, indicating that the film formed is really a mixture of ZnS and Zn hydroxides, known generically in the literature as Zn(S,OH).

**Palabras Clave:** películas delgadas, capa *buffer* de ZnS, Zn(S,OH), $Cu_2SnZnS_4$, CBD.

## 1. Introducción

La tecnología de película delgada ha hecho grandes avances en materia de fabricación de módulos fotovoltaicos de bajo costo y alta estabilidad; sin embargo, esta tecnología tiene una gran limitante ya que usa CdS en su estructura, compuesto que desde el punto de vista medio ambiental no es conveniente utilizar debido a su alta toxicidad. Por razones medioambientales, muchos centros de investigación y desarrollo en el mundo están haciendo esfuerzos para encontrar nuevos materiales que sustituyan al CdS sin deteriorar significativamente la eficiencia del dispositivo. Entre los nuevos materiales que se están investigando para sustituir el CdS, el ZnS es uno de los compuestos que ha demostrado tener propiedades ópticas y estructurales similares al CdS, convirtiéndolo en un potencial candidato a reemplazarlo; el ZnS presenta ventajas adicionales ya que la ruta de síntesis en medio acuoso es sencilla, rápida y económica y además presenta una menor toxicidad [1].

En este trabajo el compuesto ZnS se depositó usando la técnica CBD, en donde películas delgadas son crecidas sobre sustratos inmersos en soluciones que contienen iones metálicos y una fuente de calcogenuro; un agente quelante es usado para limitar la hidrólisis del ion metálico, manteniendo así su estabilidad; el no uso del agente quelante da lugar a una rápida hidrólisis que conlleva a una precipitación. Además esta técnica se basa en una lenta liberación del ion calcogenuro en la disolución y en la cual se encuentra en baja concentración el metal precursor para favorecer la formación de la película. La formación de la película delgada se lleva a cabo cuando el producto iónico (PI), excede el producto de solubilidad (PS) [2,3,4]. El mecanismo que da inicio a la formación de una película delgada por el método CBD es el proceso de nucleación, el cual ocurre tanto en fase homogénea como en fase heterogénea [5].

La nucleación en fase homogénea lleva a la formación del compuesto en forma de polvo que por gravedad se deposita en el fondo del sistema de reacción (precipitado), mientras que la nucleación en fase heterogénea permite que se forme la película sobre la superficie del sustrato; estos dos procesos pueden ocurrir simultáneamente y compiten entre sí, pero para la formación de una película se debe favorecer el proceso en fase heterogénea controlando la velocidad de crecimiento del sólido. Para que ocurra la formación de una película delgada, primero se forman centros de nucleación que se dan por adsorción de iones (mecanismo *ion by ion*) o moléculas del sólido (mecanismo *cluster-cluster*) en la superficie del sustrato, luego se da el proceso en fase heterogénea; un crecimiento a muy baja velocidad produce superficies lisas y viceversa [1,2,5].

## 2. Detalles experimentales

La síntesis de las películas delgadas de ZnS se realizó usando el método CBD; estas se depositaron sobre vidrio soda-lime y sobre vidrio recubierto con películas delgadas de $Cu_2ZnSnS_4$ depositada por co-evaporación de sus precursores metálicos en presencia de azufre [6]. Las películas se forman mediante inmersión del sustrato en una disolución con un volumen de 20 mL, constituida por acetato de zinc dihidratado ($ZnC_4H_6O_4 \cdot 2H_2O$) de concentración 22,5 mM como fuente de iones $Zn^{2+}$, citrato trisódico dihidratado ($Na_3C_6H_5O_7 \cdot 2H_2O$) 37,5 mM usado como agente acomplejante, 2,00 mL de amoniaco ($NH_3$) 0,45 % v/v para mantener el pH de la disolución (aprox. 10,5 a temperatura ambiente) y 0,360 g de tiourea ($CSN_2H_4$) como fuente de iones sulfuro $S^{2-}$; la temperatura de la solución se mantuvo siempre constante alrededor de 80 °C. Los parámetros de síntesis que se variaron para encontrar condiciones óptimas de preparación fueron: concentración de $Zn^{2+}$, concentración de tiourea y pH. Durante el estudio se mantuvo constante la temperatura de solución y el tiempo de reacción; el cuadro 1 resume el rango en que se variaron los parámetros.

| Parámetro | Rango de variación |
|---|---|
| Concentración de $Zn^{2+}$ (mM) | 15 – 60 |
| Concentración de tioúrea (mM) | 1 – 7 |
| pH | 10,15 – 10,80 |
| Tiempo (min) | 30 – 140 |

**Cuadro 1.** Rango de variación de los parámetros de síntesis.

Las películas se caracterizaron ópticamente mediante medidas de transmitancia espectral tomadas con un espectrofotómetro de alta resolución Ocean Optics HR2000CG-UV-VIS, estructuralmente mediante medidas de difracción de rayos-x utilizando un difractómetro XRD-6000 Shimadzu usando la radiación CuK$\alpha$ de un tubo polarizado a 40 KV y 30 mA, y morfológicamente con un microscopio de fuerza atómica (AFM) Autoprobe CP5 de Park Scientific Instruments. El espesor de las películas depositadas se midió con un perfilómetro Veeco Dektak 150.

## 3. Resultados y discusión

### 3.1. Caracterización estructural

Las películas delgadas de ZnS con espesores adecuados para su uso como capa buffer ($< 100$ nm) fueron inicialmente caracterizadas mediante medidas de XRD. Se encontró que películas de ZnS con espesores $< 100$ nm depositadas por CBD sobre sustrato de vidrio y sobre películas delgadas de CZTS crecen con estructura amorfa; Por el contrario, películas delgadas de ZnS de referencia depositada por co-evaporación presentaron una reflexión en $2\theta = 27,878$ que corresponde al plano (102) de la estructura hexagonal. Sin embargo el difractograma de películas depositadas por CBD con espesores mayores a 300 nm mostró reflexiones en $2\theta = 13,087$, $2\theta = 14,994$ y $2\theta = 24,836$ que corresponde a la fase $ZnSO_8H_8$ con estructura monoclínica, indicando que la película formada es realmente una mezcla de ZnS e hidróxidos de Zn a la cual genéricamente se le asigna en la literatura la fase Zn(S,OH). El difractograma del polvo precipitado en el reactor durante una deposición típica por CBD muestra las mismas reflexiones en $2\theta = 13,087$, $2\theta = 14,994$ y $2\theta = 24,836$ lo cual indica que cuando la capa depositada por CBD es muy gruesa, la mayor parte corresponde a material precipitado como resultado de la reacción homogénea.

En la figura 1 se muestran difractogramas correspondientes a películas delgadas de ZnS depositadas por co-evaporación sobre vidrio y por CBD sobre vidrio y sobre una película delgada de CZTS. Estas se comparan con el difractograma de una película delgada de CZTS depositada sobre vidrio.

Se observa que la película de ZnS sintetizada por CBD muestra un crecimiento orientado en el plano (110), distinto del ZnS depositado por co-evaporación sobre vidrio, que crece con una estructura hexagonal y presenta una reflexión en el plano (102).

El difractograma de la muestra depositado sobre vidrio recubierto por CZTS no mostró ningún pico debido a que el espesor de la película de ZnS es muy delgado (60 nm) y en este caso la capa de ZnS tiene estructura amorfa.

### 3.2. Caracterización óptica

Las Figuras 2, 3 y 4 muestran curvas de transmitancia espectral de pelícuas delgadas de ZnS depositadas sobre vidrio por CBD variando la concentración de $Zn^{2+}$, tiourea y pH. Estos resultados muestran que la transmitancia es fuertemente afectada por los parámetros de síntesis estudiados (ver recuadros). Este comportamiento se puede explicar a través de los procesos cinéticos de formación de núcleos que se llevan a cabo en fase homogénea (solución) y en fase heterogénea (interfase sustrato-solución).

La alta transmitancia observada a bajas concentraciones de $Zn^{2+}$ y tioúrea es debida fundamentalmente a un espesor pequeño de las películas de ZnS. Esto se puede explicar por una baja tasa de formación de núcleos en la superficie del sustrato, ya que la formación de núcleos tanto en la solución como en la superficie depende de las concentraciones de los precursores.

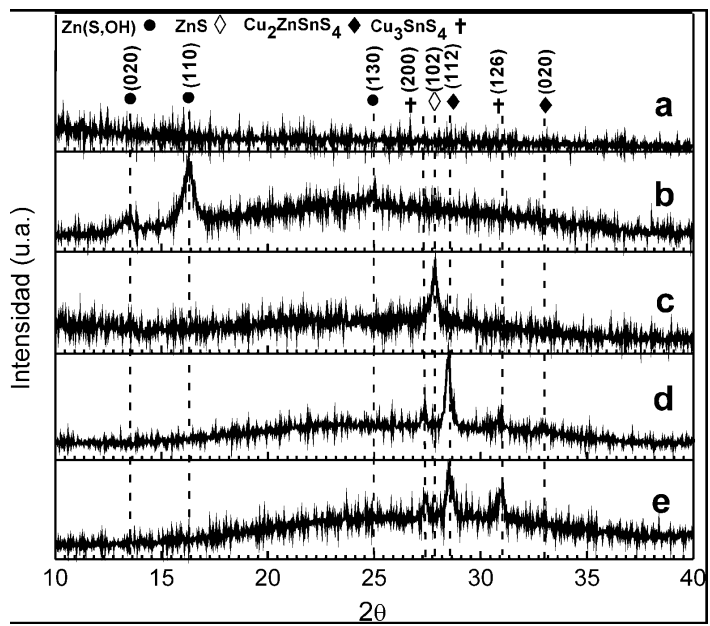

**Figura 1.** Difractogramas correspondientes a películas delgadas de: a) ZnS sintetizado por CBD sobre vidrio con espesor de 60 nm, b) ZnS sintetizado por CBD sobre vidrio con espesor del 300 nm, c) ZnS sintetizado por co-evaporación sobre vidrio con espesor 200 nm, d) $Cu_2ZnSnS_4$ depositada sobre vidrio y e) ZnS sintetizado por CBD sobre $Cu_2ZnSnS_4$ con un espesor de 60 nm.

La baja transmitancia observada con el incremento de las concentraciones de Zn y tiourea se explica principalmente por un aumento del espesor y del tamaño de grano, ocasionado por el incremento en la tasa de formación de núcleos en la superficie del sustrato al aumentar la concentración de las especies precursoras. El incremento en tamaño de grano aumenta la rugosidad superficial dando lugar a una disminución de la transmitancia.

El aumento posterior de transmitancia observado, al aumentar las concentraciones de los precursores se explica por una disminución en la velocidad de crecimiento que se presenta, pues a altas concentraciones la tasa de formación de núcleos se disminuye, debido a que el $Zn^{2+}$ no se encuentra acomplejado en su totalidad por el citrato; esto induce una mayor cantidad de $Zn^{2+}$ libre en solución y en consecuencia una menor probabilidad de formación de núcleos en la superficie.

Los resultados de la Figura 4 muestran que el pH también afecta significativamente la transmitancia. Para valores de pH menores que 10.35 la transmitancia decrece al aumentar el pH y para valores mayores de 10.35 la transmitancia aumenta, al aumentar el pH. La disminución de la trasmitancia al aumentar el pH entre 10.26 y 10.35 se explica teniendo en cuenta que en esta región el aumento

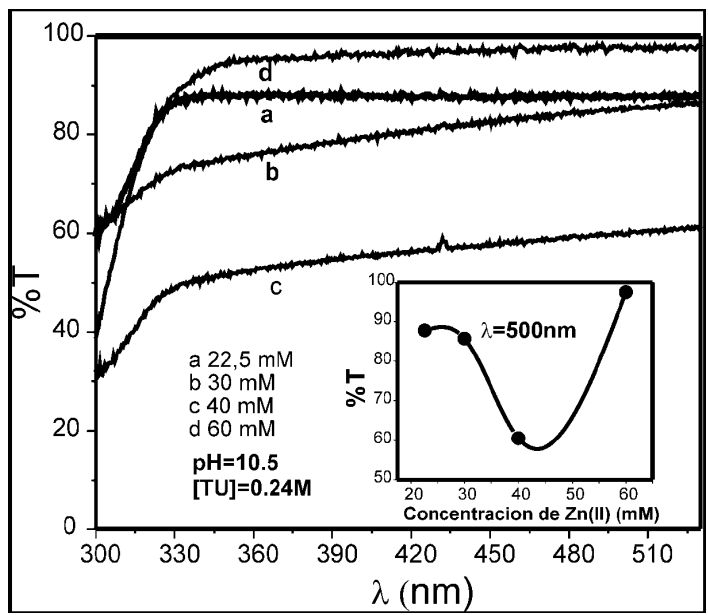

**Figura 2.** Transmitancia de películas de ZnS depositadas por CBD, variando la concentración de Zn: a)[Zn]= 22.5 mM, b) [Zn]=30 mM, c) [Zn]=40 mM y d) [Zn]=60 mM.

de pH aumenta la hidrólisis de la tiourea y se favorece la reacción heterogénea debido a que en este caso hay control cinético de la reacción:

La tiourea hidrolizada forma sulfuros que posteriormente reaccionara con el $Zn^{2+}$, dando lugar a la formación de una película de Zn(S,OH) sobre el sustrato. De esta forma la velocidad de crecimiento del Zn(S,OH) aumenta al aumentar el pH lo que da lugar a capas de mayor espesor.

De otro lado, a pH mayores que 10.35 se afecta los equilibrios químicos en los complejos que se forman en la reacción, de tal forma que al aumentar el pH aumenta la concentración de citrato el cual acomplejará una mayor proporción de $Zn^{2+}$, y por lo tanto se liberará menor cantidad de este, conllevando a una disminución de la velocidad de deposición al aumentar el pH.

En la Figura 5 se comparan curvas de variación del espesor en función del tiempo de deposición, correspondiente a películas de ZnS crecidas por el método CBD sobre sustrato de vidrio soda-lime y sobre sustrato de vidrio recubierto con una capa delgada de CZTS. Se observa que durante el crecimiento de las películas de ZnS se presentan dos procesos con diferente cinética de crecimiento; inicialmente las películas de ZnS aumentan su espesor en forma lineal y posteriormente la velocidad de crecimiento de estas tiende a saturarse. Antes de iniciar el proceso de crecimiento lineal, existe un periodo de tiempo (denominado tiempo de inducción) durante el cual tiene lugar el proceso de nucleación heterogénea que da inicio a la formación de la capa delgada. Este tiempo se determina por el cambio de color de la solución y depende principalmente de la temperatura y de la concentración de los precursores en la solución. La reducción de la velocidad de crecimiento durante la etapa de saturación es causada por una disminución de la concentración de los reactivos disponibles en la solución como consecuencia de la formación de ZnS en fase homogénea (en polvo), ya que la cantidad de reactivos consumidos para la formación de la película es mucho menor que la producida en forma de precipitado sólido.

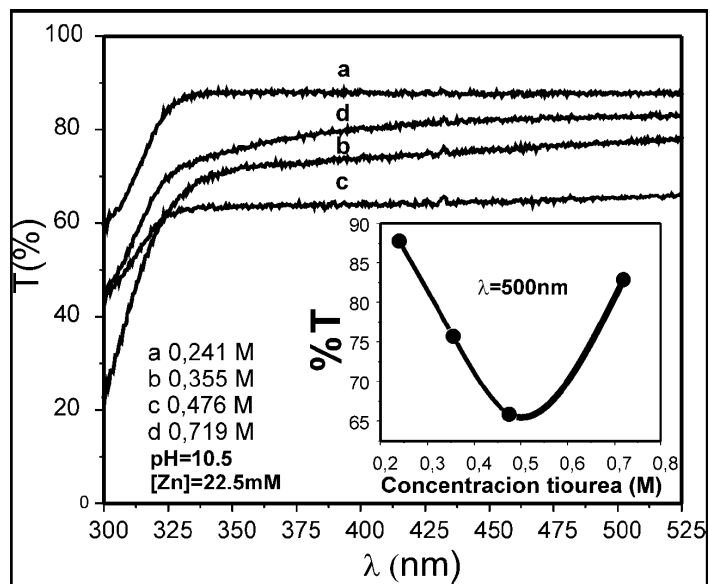

**Figura 3.** Transmitancia de películas de ZnS depositadas por CBD, variando la concentración de tiourea: a)[TU]=0.24 M, b) [TU]=0.35 M, c) [TU]=0.47 M y d) [TU]=0.71 M.

Los resultados muestran que la velocidad de crecimiento de las películas de ZnS es significativamente mayor sobre CZTS, que sobre sustrato de vidrio. La mayor velocidad de crecimiento de ZnS sobre la capa de CZTS que sobre sustratos de vidrio se explica teniendo en cuenta que el crecimiento de películas delgadas por el método CBD es afectado por los procesos de nucleación que ocurren en la superficie del sustrato; entonces, los resultados de la figura 5 indican que el sustrato de CZTS favorece el proceso de nucleación y presenta una mayor fisisorción para el proceso de crecimiento del 5 por el método CBD que la superficie del vidrio.

### 3.3. Propiedades morfológicas del ZnS

La figura 6 muestra imágenes AFM de películas delgadas de ZnS sintetizadas sobre vidrio variando la concentración de $Zn^{2+}$ entre 22.5 y 60 mM. Se observa que la concentración molar de los precursores afectan la morfología de las películas de ZnS; en particular el tamaño de grano crece con el aumento de la concentración de $Zn^{2+}$ cuando esta varía entre 22.5 y 40 mM. Posteriormente este disminuye al continuar aumentando la concentración de $Zn^{2+}$. Comparando la curva de variación del tamaño de grano en dependencia de la concentración de $Zn^{2+}$ (ver Figura 6d) con la curva de variación de la transmitancia en dependencia de la concentración de $Zn^{2+}$ (ver recuadro de la figura 2) se observa un comportamiento opuesto de estos dos parámetros con respecto a la variación de la concentración de $Zn^{2+}$, indicando que el tamaño de grano depende de la cinética de crecimiento. Velocidades de crecimiento grandes inducen tamaños de grano grandes y viceversa.

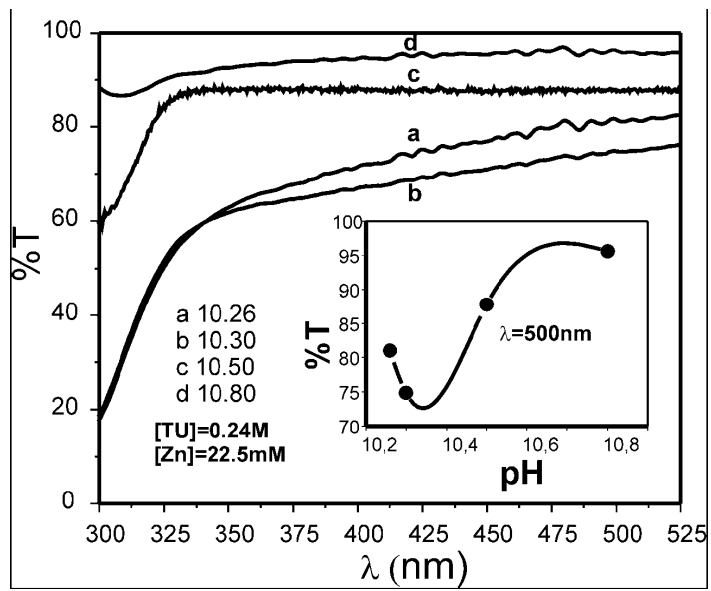

**Figura 4.** Transmitancia de películas delgadas de ZnS depositadas por CBD, variando el pH: a) 10.26 b) 10.3 c) 10.5 y d) 10.8.

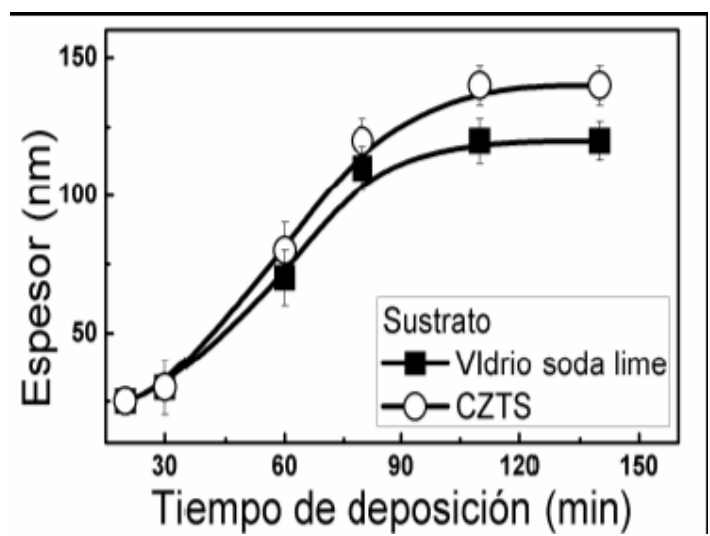

**Figura 5.** Variación del espesor vs tiempo de deposición, de películas de ZnS crecidas por CBD sobre sustrato de vidrio y sobre películas de CZTS.

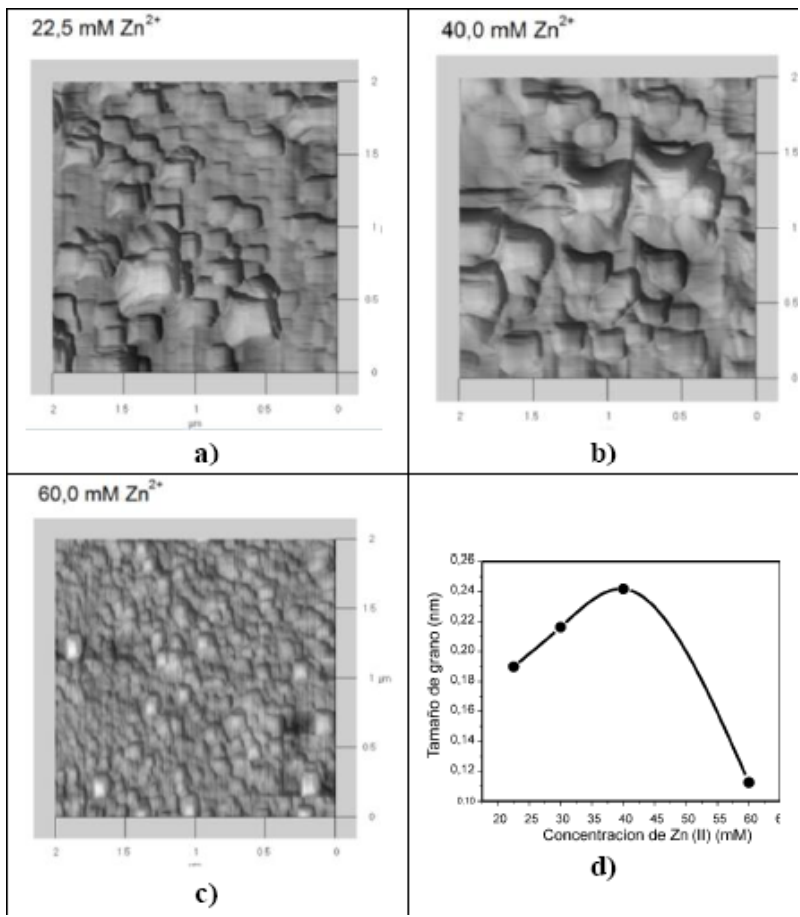

**Figura 6.** Imágenes AFM de películas delgadas de ZnS depositadas por CBD sobre vidrio a diferentes concentraciones de $Zn^{2+}$.

En la Figura 7 se compara la imagen AFM de una película delgada de CZTS depositada sobre vidrio con las imágenes AFM de películas delgadas de ZnS diferentes espesores, crecidas por CBD sobre películas delgadas de CZTS. De estos resultados se puede destacar lo siguiente:

Las películas de CZTS crecen con granos de diferente tamaño, siendo su valor promedio de 700 nm.

Las películas de ZnS depositadas por CBD sobre CZTS con espesores del orden de 400 nm (30 minutos de reacción) crecen con granos muy pequeños (120 nm en promedio) que inician su crecimiento preferencialmente en la zona de frontera de grano del CZTS. Películas de ZnS de mayor espesor (80 nm) depositadas CZTS crecen con granos de un tamaño promedio de 150 nm y la capa de ZnS formada logra cubrir la totalidad de la superficie del CZTS.

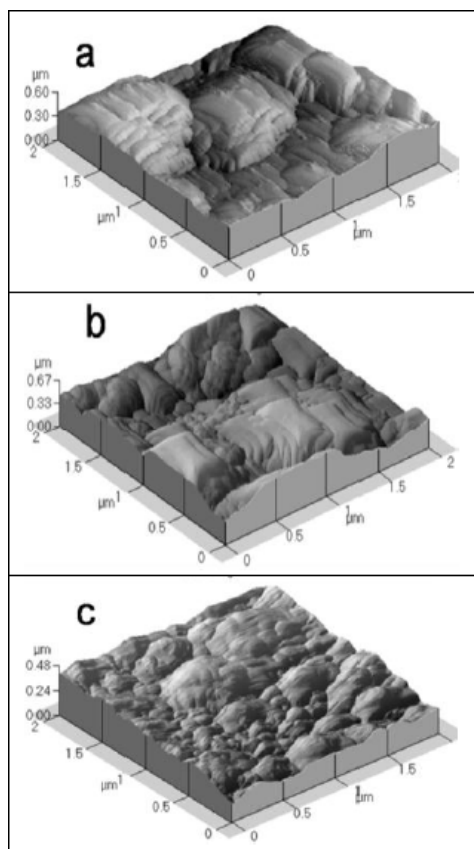

**Figura 7.** Imágenes AFM de películas delgadas de: a) CZTS depositadas sobre sustrato de vidrio b) ZnS depositada por CBD durante 30 minutos (40 nm) sobre CZTS, c) ZnS depositada por CBD durante 60 minutos (80 nm) sobre CZTS.

## 4. Conclusiones

A través de un estudio de parámetros, se encontraron condiciones para crecer en forma reproducible películas delgadas de ZnS por el método CBD usando una solución constituida por tiourea y acetato de zinc como fuentes de S y Zn y citrato de amonio como agente acomplejante. Medidas de XRD indicaron que las películas delgadas de ZnS con espesores menores que 100 nm, crecidas sobre vidrio y sobre $Cu_2SnZnS_4$ presentan una estructura amorfa; sin embargo, muestras con espesores mayores de 300 nm crecen con estructura cristalina correspondiente a la fase $ZnSO_8H_8$ (sulfohidróxido de zinc), indicando que la película formada es realmente una mezcla de ZnS e hidróxidos de Zn.

La transmitancia de las películas de ZnS es afectada principalmente por el espesor y el tamaño de grano, los cuales dependen a su vez de procesos cinéticos de formación de núcleos que se llevan a cabo en fase homogénea y en fase heterogénea; estos procesos son críticamente afectados por la concentración de precursores y pH.

La cinética de crecimiento de las películas de ZnS es afectada por el tipo de sustrato. Estas crecen más rápidamente sobre sustrato de CZTS que sobre vidrio, indicando que el primero favorece la nucleación más efectivamente que el segundo.

## Referencias